\documentclass{aa}
\usepackage{graphicx}
\usepackage{natbib}
\begin{document}
   \title{High-energy neutrino as observational signature \\
of massive black hole formation}
   \author{V. Berezinsky
          \inst{1,2}
          \and
          V. Dokuchaev
          \inst{2}
          }
   \offprints{V. Dokuchaev}
   \institute{Laboratori Nazionali del Gran Sasso, INFN,
67010 Assergi (AQ), Italy \\
   \email{berezinsky@lngs.infn.it}
   \and
Institute for Nuclear Research of the Russian Academy of Sciences, \\
60th October Anniversary prospect 7a, 117312 Moscow, Russia \\
    \email{dokuchaev@inr.npd.ac.ru}
              }
\date{Received November 8, 2004; accepted XXX XX, XXXX}

\abstract{We describe the formation of a seed massive black hole (MBH)
inside a supermassive star (SMS) in a distant galactic nucleus. The
short-lived SMS is naturally formed  due to collision destructions of
normal stars in the evolving galactic nucleus. The neutron stars (NSs)
and stellar-mass black holes form a compact self-gravitating subsystem
deep inside a SMS. This subsystem is short-lived in comparison with a
host SMS and collapses finally into the MBH. Just before gravitational
collapse of compact subsystem the frequent NS collisions are
accompanied by the generation of numerous ultra-relativistic fireballs.
A combined ram pressure of multiple coexisting fireballs produces a
quasi-stationary rarefied cavity in the central part of SMS. The
protons are accelerated in the fireballs and by  relativistic shocks in
the cavity. All secondary particles, produced in collisions, except the
high-energy neutrinos are absorbed in the SMS interiors.  An estimated
high-energy neutrino signal from this hidden source can be detected by
the neutrino telescope with an effective area $S\sim1$~km$^2$ providing
the evidence for MBH formation in a distant galactic nucleus. A
corresponding lifetime of this high-energy hidden neutrino source is
$\sim 0.1-1$~yr.
       }

   \titlerunning{High-energy neutrino}

   \maketitle

\section{Introduction}

The origin of massive black holes (MBHs) in the galactic nuclei is a
long standing problem. In this paper we describe a `natural' scenario
of a seed MBH formation in the process of dynamical evolution of a
central stellar cluster in the galactic nuclei. We demonstrate that
high-energy neutrinos may be the observational signature of MBH
formation.

High-energy (HE) neutrino radiation from astrophysical sources is
accompanied by other types of radiation, most notably by the HE
gamma-radiation. This HE gamma-radiation put an upper limit on the
neutrino flux emitted by the neutrino-transparent astrophysical
sources. For example, if neutrinos are generated by the interaction of
HE protons with low energy photons in the extragalactic space or in
gamma-transparent sources, the upper limit on diffuse neutrino flux
$I_{\nu}(E)$ can be derived from electro-magnetic cascade radiation
\citep{BeSm}. However, there can be the ``hidden''
neutrino sources where accompanying X-ray and gamma-radiations are
absorbed. There are several known models of the hidden HE
neutrino sources: {\em The Young SN shells} \citep{BePr}, opaque for
all radiation except of neutrino during a time interval
$t_{\nu}\sim10^3-10^4~$s. {\em The Thorne-Zytkow star} \citep{ThZy},
which is a neutron star submerged into a red giant.
{\em The cocooned massive
black hole} in AGN \citep{ber81} where the electromagnetic radiation
is absorbed in a cocoon around the black hole. {\em The AGN with a
standing shock} in the vicinity of MBH \citep{Ste91} can produce large
flux of HE neutrinos accompanied by the relatively small flux of X-ray
radiation.

Recently we proposed the model for a powerful but short-lived hidden
neutrino source which originates in a distant galactic nucleus prior to
its collapse into the MBH \citep{ber01}. In this model we assume that
MBH is formed by the natural dynamical evolution of the central stellar
cluster in the normal galactic nucleus. Dynamical evolution of dense
central stellar clusters in the galactic nuclei is accompanied by the
growth of a central gravitational potential $\phi\sim v^2$, where  $v$
is a velocity dispersion of constituent stars. This evolution process
is accompanied by the contraction of a central part of the galactic
nucleus and terminated by formation of MBH (see for a review
\cite{ree84} and references therein). During this evolution a dense
galactic nucleus inevitably proceeds through the stellar collision
phase  \citep{spi66,spi71,ree84,dok91}, when normal stars in the
cluster are destroyed in mutual collisions. A direct collision of two
stars ends with their complete destruction if their relative kinetic
energy exceeds the gravitational bound energy of stars. This condition
is realized in stellar clusters with a high enough velocity dispersion.
Namely, $v\geq v_{\rm esc}=(2Gm_*/r_*)^{1/2}$ is fulfilled,  where
$v_{\rm esc}$ is an escape velocity from the surface of star with mass
$m_*$ and radius $r_*$. For a solar type star $v_{\rm
esc}\simeq620$~km~s$^{-1}$. In a stellar cluster with $v>v_{\rm esc}$
the normal stars are eventually destroyed in mutual collisions or in
collisions with the extremely compact stellar remnants: neutron stars
(NSs) or stellar mass black holes. Only compact stellar remnants
survive through the normal star-destruction stage of evolution at
$v\simeq v_{\rm esc}$ and form the self-gravitating subsystem
surrounded by a massive gas envelope. We shall refer to all compact
stellar remnants as the NSs for brevity. The fireballs after numerous
NS collisions in the contracting NS cluster result in the formation of
a rarefied cavity in the massive gas envelope. Particles accelerated in
this rarefied cavity interact with a dense envelope and produce
secondary HE neutrinos. The accompanying gamma-radiation is fully
absorbed in the case of thick enough envelope. The resulting hidden HE
neutrino source is short-lived and very powerful: neutrino luminosity
may exceed the Eddington limit for the electromagnetic radiation.

\cite{ber01} considered the case when a self-gravitating
cluster of NSs in the evolving galactic nucleus is formed {\em
simultaneously} with a formation of massive gas envelope. It is
possible because the time of collision of self-destruction for normal
stars appears to be of the same order as the time of NS cluster
formation at a star-destruction stage of evolution of the galactic
nucleus (see Section~\ref{NScluster} for more details). In this case
the newly formed NS cluster evolves much faster than a surrounding
massive envelope. The corresponding parameters of massive envelope are
fixed at the moment of NS cluster formation when $v\simeq v_{\rm esc}$.
Here we will consider another and probably more natural case, when a
self-destruction of normal stars and formation of a massive envelope
in the evolving galactic nucleus proceeds faster than a formation of
the self-gravitating subsystem of NSs in it. This case corresponds to
the initial formation of a supermassive star (SMS) with the individual
NSs submerged in it. Correspondingly, a formation of self-gravitating
subsystem of NSs is {\em delayed} with respect to the formation of
SMS, and the evolutionary contraction of SMS must be taken into
account.  We show that the hidden source of HE neutrino arises in this
case also and even more powerful than in the first scenario
\citep{ber01} of a simultaneous formation of a massive envelope and NS
cluster in the evolving galactic nucleus.

\section{Supermassive star formation}
\label{SMS}

The SMS may be formed from the gas produced in the process of
destructive collisions of stars in the evolved stellar cluster with a
velocity dispersion $v\geq v_{\rm esc}$. A characteristic time-scale
for stellar cluster dynamical evolution is the (two-body) relaxation
time
\begin{equation}
 \label{tr}
 t_{\rm r}=\left(\frac{2}{3}\right)^{\!1/2}\!\!\!
 \frac{v^3}{4\pi G^2m^2n\Lambda}
 \simeq4.6\times10^8N_8^2\left(\frac{v}{v_{\rm esc}}\right)^{-3}
 \!\mbox{ yr},
\end{equation}
where $N=10^8N_8$ is the number of stars in the cluster,
$\Lambda=\ln(0.4N)$ is the Coulomb logarithm, $v$ is a star velocity
dispersion, $n$ is a star number density, $m\simeq{\rm M}_{\odot}$ is a
mass of constituent stars. The corresponding virial radius of the
stellar cluster is $R=GNm/2v^2$. The first equality in (\ref{tr}) is
valid for the local values of parameters, meanwhile, the second one is
valid only for the mean (virial) parameters of a self-gravitating
cluster. At $v>v_{\rm esc}$, where $v_{\rm esc}$ is an escape velocity
from the surface of constituent star, the time-scale for
self-destruction of normal stars in mutual collisions is $t_{\rm
coll}=(v_{\rm esc}/v)^4\Lambda t_{\rm r}$ \citep[e.~g.][]{dok91}.
Numerical modelling of catastrophic stellar collisions has been
performed by e.~g. \cite{ben92,lai93}. We choose $v\simeq v_{\rm esc}$
as a characteristic threshold value for a complete destruction of two
stars and final production of unbound gas cloud. As a result, the
stellar cluster in the evolved galactic nucleus with $v\geq v_{\rm
esc}$ transforms finally into the SMS due to catastrophic stellar
collisions. At $v\simeq v_{\rm esc}$ the time-scale for the formation
of SMS due to destructive collisions of stars is of the same order as
the relaxation time, $t_{\rm coll}(v_{\rm esc})\sim t_{\rm r}(v_{\rm
esc})$. A total mass of the gas produced by destruction of normal stars
composes the major part of a progenitor central stellar cluster in the
galactic nucleus. Thus, the natural range of masses for the formed SMS
is $M_{\rm SMS}=10^7-10^8{\rm M_{\odot}}$.

A newly formed SMS with mass $M_{\rm SMS}$ and radius $R_{\rm SMS}$
gradually contracts due to radiation with the Eddington luminosity
$L_{\rm E}=4\pi GM_{\rm SMS} m_p c/\sigma_{\rm T}$, where $m_p$ is the
proton mass and $\sigma_{\rm T}$ is the Thompson cross-section. A
nonrotating SMS is a short-lived object that collapses due to
post-Newtonian instability. Rotation provides the stabilization of SMS
if the rotation energy is an appreciable part of its total energy
$E_{\rm SMS}\simeq(GM_{\rm SMS}^2/2R_{\rm SMS})$. In general an
evolution time of SMS is the Kelvin-Helmholtz time-scale $t_{\rm SMS}=
E_{\rm SMS}/L_{\rm E}\propto R_{\rm SMS}^{-1}$. Stabilization of SMS by
rotation (and additionally by internal magnetic field) ensures in
principle its gradual contraction up to the gravitational radius
\citep{zel71,sha83,new01}. Thus, the maximum evolution time of SMS is
of the order of the Eddington time $t_{\rm E}=0.1M_{\rm SMS}c^2/L_{\rm
E}\simeq4.5\times10^7$~yr. We approximate the distribution of gas in a
SMS by polytropic model with an adiabatic index  $\gamma=4/3$. For this
value of adiabatic index the central gas density in SMS is
$\rho_{c}=k_{\rm c} n_{\rm SMS}m_{\rm p}$ and central sound velocity
$c_{\rm s,c}=k_{\rm s} v_{\rm SMS}$, where $n_{\rm SMS}$ is a SMS mean
number density, $v_{\rm SMS}= (GM_{\rm SMS}/2R_{\rm SMS})^{1/2}$ is a
SMS virial velocity and numerical constants $k_{\rm c}\simeq54.2$ and
$k_{\rm s}\simeq1.51$ respectively.

\section{Neutron star cluster formation}
\label{NScluster}

The compact stellar remnants in the form of NSs and stellar mass black
holes submerged into the newly formed SMS. This is a specific feature
of SMS formed in the evolving stellar cluster. Let  SMS contains a
small fraction of identical NSs with a total mass $M_{\rm NS}=f_{\rm
NS}M_{\rm SMS}$, $f_{\rm NS}\ll1$ and a typical mass of individual NS
is $m_{\rm NS}=1.4{\rm M_{\odot}}$. In numerical estimations we will use
$f_{\rm NS}=10^{-2}f_{-2}$ with $f_{-2}\sim1$. An individual NS with a
mass $m$ and local velocity $V$ is spiraling down the SMS center under
influence of the dynamical friction force  \citep{cha43,ost99} $F_{\rm
df}=4\pi I(Gm)^2\rho_{\rm SMS}/V^2$, where $\rho_{\rm SMS}$ is a SMS
gas density. The dimensionless factor here is $I\simeq 1$ in the case of
$V\simeq c_{\rm s}$, where $c_{\rm s}$ is a sound velocity in SMS.
Deep inside SMS the individual NSs are spiralling down and combine into a
fast evolving self-gravitating subsystem (NS cluster). A corresponding
time-scale for the NS dynamical friction drag toward the SMS center is
$t_{\rm df}=V/\dot V=mV/F_{\rm df}$. At the moment of SMS formation,
when $V\sim c_{\rm s}\sim v_{\rm esc}$, a time-scale of NS friction
drag is of the same order as the time-scale of SMS
formation, $t_{\rm df}(v_{\rm esc})\sim t_{\rm coll}(v_{\rm esc})\sim
t_{\rm r}(v_{\rm esc})$. Thus, in general it is possible either (i) {\em
a simultaneous formation} of a self-gravitating cluster of NSs and the
host SMS or (ii) {\em a delayed formation} of a self-gravitating
cluster of NSs inside the host SMS. The case (i) was considered by
\cite{ber01} and the case (ii) we describe below. In particular, in
the case (ii) it must be taken into account a possibility of the SMS
contraction prior to the formation of a central NS cluster.

The NS dynamical friction time-scale for a mean NS inside the host SMS
is {\em decreasing} during contraction of SMS as
$t_{\rm df}\propto R^{3/2}$. On the contrary, the SMS evolution
time-scale is {\em increasing} with SMS contraction as $t_{\rm
SMS}\propto R^{-1}$. Thus, the subsystem of NSs begins evolving faster
than the host SMS after reaching the stage when $t_{\rm df} \simeq
t_{\rm SMS}$. At this stage all NSs sink deep to a central part of SMS
and form there a self-gravitating cluster. From relation $t_{\rm
df}(R) \simeq t_{\rm SMS}(R)$ we find the corresponding SMS radius at
the moment of NS cluster formation inside it:
\begin{equation}\label{Rsms}
  R_{\rm SMS} \simeq\!\!\left(\frac{9I^2}{8\pi^2}
 \frac{G m_{\rm NS}^2M_{\rm SMS}
  \sigma_{\rm T}^2}{c ^2 m_{\rm  p}^2}\right)^{\!\!1/5}\!\!\!\!
  \simeq4.6\times10^{15}M_8^{1/5} \!\mbox{ cm},
\end{equation}
where $M_8=M_{\rm SMS}/(10^8{\rm M_{\odot}})$. Respectively a mean gas
density
\begin{equation}
 \label{rhosms}
  n_{\rm SMS}
  \simeq \frac{3}{4\pi}\frac{M_{\rm SMS}}{R_{\rm SMS}^3
  m_{\rm p}} \simeq2.9\times10^{17}M_8^{2/5}\mbox{ cm}^{-3},
\end{equation}
a mean column density
\begin{equation} \label{depth}
  X_{\rm SMS}\simeq n_{\rm SMS}m_{\rm p} R_{\rm SMS}
  \simeq2.2\times10^9M_8^{3/5} \mbox{ g cm}^{-2},
\end{equation}
a virial velocity $v_{\rm SMS}= (GM_{\rm SMS}/2R_{\rm SMS})^{1/2}
\simeq 0.04M_8^{2/5}c$, a SMS evolution time $t_{\rm SMS} = t_{\rm
SMS}(R_{\rm SMS}) \simeq 7.3 \times10^5 M_8^{4/5}$~yr and a surface
black-body temperature $T_{\rm SMS}\simeq 3\times10^4M_8^{3/20}$~K.
Note that the only free parameter in these relations is $M_{\rm SMS}$.
A self-gravitating NS cluster with a radius $R_{\rm NS}\ll R_{\rm SMS}$
is formed when a total mass of NSs inside this radius becomes of order
of an ambient gas mass, $M_{\rm NS}\simeq(4\pi/3)R_{\rm NS}^3\rho_{\rm
c}$, where $\rho_{\rm c}$ is a central SMS density. From this relation
the initial radius of NS cluster is $R_{\rm NS}=f_{\rm NS}^{1/3}k_{\rm
c}^{-1/3}R_{\rm SMS}$ and velocity dispersion $v_{\rm NS}=f_{\rm
NS}^{1/3}k_{\rm c}^{1/3}v_{\rm SMS}$, respectively.

\section{Neutron star cluster evolution}

The formed NS cluster evolves much more faster than a host SMS and
collapses finally into the MBH. A suitable approximation for the
dynamical evolution of NS cluster in our case is the homologous
``evaporation'' model \citep[e.~g.][]{Spitzer, Saslaw}. This model
describes the two-body interactions of stars in the cluster with an
assumption that a fast star escapes (evaporates) from the cluster with
a zero total energy. Accordingly, a total virial energy of the cluster
remains constant $E=-Nmv^2/2=const$ during evaporation of fast stars.
The  velocity dispersion in the evolving cluster is growing as $v
\propto N^{-1/2}$ and cluster radius diminishes as $R\propto N^2$
with a diminishing of number $N$ of remaining stars in the cluster.  After
reaching $v\simeq0.3c$, which is onset of the NS cluster global
dynamical instability, the remaining NS cluster collapses to the MBH
\citep[e.~g.][]{zp65,qui87}). The rate of NS evaporation from the
cluster is
\begin{equation}
  \dot N_{\rm ev} \simeq \alpha N t_{\rm r}^{-1}, \label{Nev}
\end{equation}
where a relaxation time $t_{\rm r}$ is given by (\ref{tr}) and constant
$\alpha \sim10^{-3} - 10^{-2}$ according to the  numerical calculations
in the Fokker-Plank approximation \citep{Spitzer, Saslaw,coh80}.
Integration of (\ref{Nev}) together with the  relation $E=const$
results in the nondissipative evolution due to evaporation of fast
stars:
\begin{equation}
  N(t)=N_{\rm NS}\left(1-\frac{t}{t_{\rm ev}}\right)^{2/7},
  \label{Nt}
\end{equation}
where the cluster evolution time $t_{\rm ev}=a_{\rm ev}t_{\rm r}$,
numerical constant $a_{\rm ev}=(2/7)\alpha^{-1}$ and $t_{\rm r}$ is the
relaxation time at the moment of NS cluster formation $t=0$. In
numerical estimations we put $a_{\rm ev}=10^2a_{2}$ with $a_{2}\sim1$.

The random collisions of NSs and their evaporation from the cluster
occur at the same time. A rate of NS collisions in the cluster with the
gravitational radiation losses taken into account \citep{qui87,dok98}
is
\begin{eqnarray}
 \dot N_{\rm cap}  \simeq
 36\sqrt2\left(\frac{v}{c}\right)^{31/7}\frac{1}{N}\frac {c}{r_{\rm
 g}},
 \label{Ncrate}
\end{eqnarray}
where $r_{\rm g}=2Gm_{\rm NS}/c^2$ is a gravitational radius of NS.
According to (\ref{Nt}) the collision rate (\ref{Ncrate}) is gradually
growing due to the evaporative dynamical evolution of cluster as $\dot
N_{\rm cap}\propto v^{45/7}$. The dissipative process of collisions is
negligible for the dynamical evolution of a cluster in comparison with
the nondissipative evaporation until $\dot N_{\rm cap}<\dot N_{\rm
ev}$. This inequality is equivalent to the condition $v<v_{\rm cap}$,
where
\begin{equation}
  v_{\rm cap}=
  \left(\frac{4}{7\sqrt3}\frac{\Lambda}{a_{\rm ev}}\right)^{7/10}c
  \simeq8.7\times10^{-2}a_{2}^{-7/10}c\,.
  \label{vcap}
\end{equation}
At $v>v_{\rm cap}$ the NS collisions becomes a major evolution process
in cluster. The contraction of cluster is accelerated and it finally
collapses to MBH.

\section{Fireballs and rarefied cavity inside the SMS}
\label{cavity}

We assume that each NS collision is accompanied by the generation of
relativistic fireball with a total energy $E_0=10^{52}E_{52}$~erg,
where $E_{52}\sim1$. The physics of fireballs is extensively elaborated
in the recent years for the gamma-ray burst (GRB) models
\citep[for review see e.~g.][ and references therein]{pir00}.

Although in our case at least the first fireball is born in SMS
with large density $n_{\rm SMS}$ given by Eq.~(\ref{rhosms}),
the dynamics of the
fireballs remains the same as in the standard theory. Consider
the first fireball assuming it to be spherically
symmetric, which in fact is not a necessary assumption.

The tremendous energy release $E_0 \sim 10^{52}$~erg occurs
within the small volume with radius $R_i \sim 10^7$~cm and the
radiation pressure $p_i \sim E_0/R_i^3$ by many orders of magnitude
exceeds the external pressure $p \sim n_{SMS}kT$.
The pressure in the fireball, which
consists of relativistic $\gamma e^+e^-$ gas with small admixture of
the baryons, accelerates fireball to large Lorentz-factor $\Gamma_f$.
The relativistic shock, with Lorentz factor
$\Gamma_{\rm sh}=\sqrt{2} \Gamma_f$ propagates ahead of the fireball
pushing the SMS gas. The shock becomes non-relativistic at the Sedov
length
\begin{equation}
l_s =(3E_0/4\pi n_{\rm SMS}m_Hc^2)^{1/3} =1.7 \times 10^{12}~{\rm cm}.
\label{L_s}
\end{equation}
This value gives an order of magnitude estimate for cavity radius (see
below).

The second fireball propagates in the rarefied cavity produced by the
first fireball, and it overtakes the first one at about Sedov length,
when the first shock is non-relativistic. The shocks from the first
and successive fireballs merge into one, hence the following
scenario is plausible.

The numerous fireballs after successive NS collisions in a dense cluster
provide the central energy source in our model. The power of this
source is not limited by the Eddington luminosity in contrast to the
accretion sources. The radius of NS cluster is very small in comparison
with the host SMS radius $R_{\rm SMS}$. The shocks
produced by numerous fireballs dig out a rarefied cavity with a small
radius in comparison with $R_{\rm SMS}$. Consider the merging shock
in more details.

The early {\em nonstationary} stage of shock expansion can be described by
the self-similar spherical shock solution for a central source with energy
varying in time as $E=A t^k$ with $A=const$ and $k=const$
\citep{ost88,dok02}. The particular case of $k=0$ corresponds to the
Sedov \emph{instant shock} solution \citep[see e.~g.][]{ll59}.
Meanwhile the nonstationary shock from multiple fireballs corresponds
to a permanent energy injection into the shock, or \emph{injection
shock} with a central source of constant luminosity, $k=1$, $L=A$,
$E=L t$. The radius of a self-similar expanding shock grows with time
as
 \begin{equation}
 R= R(t)=\beta\left(\frac{A}{\rho_{\rm c}}\right)^{1/5}t^{(2+k)/5},
 \label{Rt}
 \end{equation}
where numerical constant $\beta=\beta(\gamma,k)$ depends on the gas
adiabatic index $\gamma$ and $k$, e.~g. $\beta(4/3,1)=0.793$
\citep{dok02}. The velocity of shock expansion is
\begin{equation}
 u= \frac{dR}{dt}=\frac{(2+k)R}{5t}\,.
 \label{u1}
\end{equation}
The maximum radius of the expanding strong shock $R_{\rm sh}$ is
obtained from the equality $u(R_{\rm sh}) = c_{\rm s,c}$ by using
(\ref{Rt}) and (\ref{u1}):
\begin{equation}
 R_{\rm sh}=
 \left[\left(\frac{2+k}{5\,c_{\rm s,c}}\right)^{2+k}\beta^{5}\,
 \frac{A}{\rho_{\rm c}}\right]^{1/(3-k)}.
 \label{Rsh}
\end{equation}
Respectively the maximum time of a strong shock expansion is $t_{\rm
sh} = [(2+k)/5](R_{\rm sh}/c_{\rm s,c})$. The expansion law for a $k=1$
injection shock corresponds to \emph{constant energy flux} (or constant
source luminosity) carried by the swept-out gas. This constant
luminosity shock solution is applicable only to the early nonstationary
stage of cavity formation at $t<t_{\rm sh}$.

At the late {\em stationary} stage at $t\geq t_{\rm sh}$ the boundary
of the cavity is supported in a dynamic equilibrium by the
relativistic wind from successive fireballs. The radius of a
stationary cavity $R_{\rm cav}$ is determined from the energy flux
balance on its boundary. The central source power or luminosity is
$L=\dot N_{\rm cap} E_0$, where $\dot N_{\rm cap}$ is from
(\ref{Ncrate}). After formation of cavity the energy generated in the
central source is carried out by a relativistic wind. Just outside
$R_{\rm cav}$ the energy flux is carried by a hydrodynamic flow
$L=4\pi R_{\rm cav}^2 \rho v(w+v^2/2)$ with a gas velocity $v \sim
c_{\rm s,c}$, where $c_{\rm s,c}$ and $\rho_{\rm c}$ are respectively
a sound velocity and gas density in the central part of SMS. The gas
produces some work under expansion, so the energy flux $L$ contains an
enthalpy density $w=\varepsilon+p/\rho=c_{\rm s}^2/(\gamma-1)$, where
$\varepsilon= c_{\rm s}^2/[\gamma(\gamma-1)]$, $p$, $\rho$ and
$\gamma$ are respectively a gas internal energy density, pressure,
density and adiabatic index. From the energy flux balance relation
we determine a stationary cavity radius for $\gamma=4/3$
\begin{equation}
 \label{Rcav}
 R_{\rm cav}=\left(\frac{\dot N_{\rm cap}E_0}
 {14\pi\rho_{\rm c}c_{\rm s,c}^3}\right)^{1/2}.
\end{equation}
For the stationary cavity existence the time interval between successive
fireballs $t_{\rm cap}=\dot N_{\rm cap}^{-1}$ must be less than that for
cavity shrinking (or spreading) $t_{\rm cav}=R_{\rm cav}/c_{\rm
s,c}$. This requirement, $t_{\rm cap}<t_{\rm cav}$, with a help of
(\ref{Ncrate}) and (\ref{Rcav}) can be written as condition $v>v_{\rm
cav}$, where
\begin{equation}
 \label{vcav}
  v_{\rm cav}\simeq
  7.3\times10^{-2}f_{-2}^{7/27}E_{52}^{-7/135}M_8^{91/225}c\,,
\end{equation}
with numerical values $v_{\rm cav}\simeq4.3v_{\rm NS}$. The
corresponding minimal radius of a stationary cavity $R_{\rm min}=R_{\rm
cav}(v_{\rm cav})$ is
\begin{equation}\label{Rmin}
  R_{\rm min}\simeq
  \left(\frac{E_0}{4\pi\rho_{\rm c}c_{\rm s,c}^2}\right)^{1/3}
  \simeq2.1\times10^{12}E_{52}^{1/3}M_8^{-2/5} \mbox{ cm}.
\end{equation}
It follows from (\ref{Rmin}) that $R_{\rm min}$ is independent of
$f_{\rm NS}$. At the moment of a stationary cavity formation inside
SMS, $v=v_{\rm cav}$, we have respectively a radius of the  NS cluster
$R(v_{\rm cav}) \simeq 7.5\times10^{11}
f_{-2}^{17/27}E_{52}^{28/135}M_8^{41/225}$~cm, a number of NSs in the
cluster at the moment of cavity formation $N(v_{\rm cav}) \simeq
3.8\times10^4 f_{-2}^{31/27}E_{52}^{14/135}M_8^{223/225}$, a NS cluster
evolution time $t_{\rm ev}(v_{\rm cav}) \simeq
0.3a_{2}f_{-2}^{41/27}E_{52}^{49/135}M_8^{173/225}$~yr and central
source luminosity inside a SMS $L(v_{\rm cav}) \simeq 8.6\times10^{48}
E_{52}^{2/3}M_8^{4/5}$~erg/s. After the formation of a stationary
cavity inside the SMS we have a hierarchy of radii, $R(v_{\rm cav}) <
R_{\rm min} \ll R_{\rm SMS}$, which justifies the using of central
values for a gas density $\rho_{\rm c}$ and sound velocity $v_{\rm
s,c}$.

The central source power and cavity radius are gradually growing due to
the evolutionary grows of NS collision rate. The rate of NS collisions
and the power of the central source reach the maximum value at $v\simeq
v_{\rm cap}$ from (\ref{vcap}). The corresponding maximum central
source power $L_{\rm max}=\dot N_{\rm cap}(v_{\rm cap})E_0$ is
\begin{equation} \label{Lmax}
  L_{\rm max}\simeq2.8\times10^{49}
  a_2^{-9/2}f_{-2}^{-5/3}E_{52}M_8^{-9/5}\mbox{ erg~s}^{-1}.
  \end{equation}
By substituting this luminosity in (\ref{Rcav}) we find that the
maximal cavity radius $R_{\rm max}=R_{\rm cav}(v_{\rm cap})$ is
\begin{equation} \label{Rmax}
  R_{\rm max}\simeq3.8\times10^{12}
  a_{2}^{-9/4}f_{-2}^{-5/6}E_{52}^{1/2}M_8^{-17/10} \mbox{ cm}.
\end{equation}
Respectively at $v=v_{\rm cap}$ the NS cluster radius $R(v_{\rm
cap})\simeq 3.6\times10^{11} a_{2}^{14/5}f_{-2}^{5/3} M_8^{9/5}$~cm,
number of NS in the cluster $N(v_{\rm cap}) \simeq
2.7\times10^4a_{2}^{7/5} f_{-2}^{5/3}M_8^{9/5}$ and NS cluster
evolution time $t_{\rm ev}(v_{\rm cap}) \simeq
31.3a_{2}^{59/10}f_{-2}^{10/3}M_8^{18/5}$~days.

Collisions of NSs during a cluster lifetime $t_{\rm ev}$ at $v=v_{\rm
cap}$ supply the total energy $\mathcal{E}_{\rm max}\simeq L_{\rm max}
t_{\rm ev} \simeq 7.6\times
10^{55}a_2^{7/5}f_{-2}^{5/3}E_0M_8^{9/5}$~erg into the cavity. This
energy is far less than a SMS binding energy $E_{\rm SMS}\simeq GM_{\rm
SMS}^2/2R_{\rm SMS} \simeq2.9\times10^{59}M_8^{9/5}$~erg, and so the
shocks after numerous fireballs do not influence the SMS internal
structure.

\section{Production and detection of high-energy neutrinos}

We shall study here the acceleration of protons by relativistic
fireballs produced by NS collisions and the high-energy neutrino
radiation produced in $p\gamma$ and $pp$ collisions. We assume a
standard GRB fireball which propagates at the baryon-dominated stage
with the Lorentz-factor $\Gamma= 10^2\Gamma_2$, $\Gamma_2\sim1$ and
carries a baryonic mass $M_0= E_0/\Gamma c^2\simeq
5.6\times10^{-5}E_{52}\Gamma_2^{-1}{\rm M_{\odot}}$.

The acceleration occurs due to internal shocks inside the fireball,
\cite{wax95}, and due to external shocks, \cite{Vietri95}. We shall
consider first the internal shock acceleration and high-energy neutrino
production following \cite{wax95,wax01}. This scenario we consider as
the basic one for a prediction of neutrino flux in our model.

The standard explanation of the observed GRBs, \cite{pir00}, is given
under an assumption that fireball consists of the mildly relativistic
sub-shells. The inner sub-shells are moving faster than the outer ones,
and their collisions produce mildly relativistic shocks. The protons
are accelerated by these shocks with the standard spectrum $\propto
E^{-2}$ and with a maximum acceleration energy $E_{\rm max} \sim
1\times 10^{21}$~eV, \cite{wax95}. The specific feature of our case is
a relatively short time of fireball existence in the cavity, $t \sim
R_{\rm cav}/c$. This time, however, is long enough for acceleration of
protons up to the maximum energy $E_{\rm max} \sim 10^{21}$~eV. In the
rest-frame ${\cal K'}$ the acceleration time is $t_a' \sim r'_L/c$,
where $r'_L$ is the Larmor radius. Following to \cite{pir00}, an
equipartition magnetic field $B'_{\rm eq}$ in the relativistic wind can
be estimated as
\begin{equation}
 \frac {B'^2_{\rm eq}}{8\pi} \sim \xi_B \omega'_{\gamma} \sim
 2\xi_B\frac{L_{\gamma}} {\Gamma^2 r^2 c}\,,
 \label{B_eq}
\end{equation}
where unprimed symbols correspond to quantities in the laboratory
frame, $\omega'_{\gamma}$ is photon energy density, $\xi_B \sim 1$ is
an equipartition parameter and $L$ is a fireball luminosity. Using
$L_{\gamma} \sim 10^{52}$~erg/s and $r \sim R_{\rm cav}$ we obtain the
acceleration time as $t_a \sim 50$~s for particles with $E_{\rm max}
\sim 10^{21}$~eV. With an equipartition magnetic field (\ref{B_eq}) the
maximum acceleration energy is $E \sim eB'_{\rm eq}R_{\rm cav} \sim
10^{21}$~eV in agreement with the \cite{wax95} model.

High-energy neutrinos are produced in $p\gamma$ collisions inside the
fireball by the \cite{WaBa97} mechanism. Due to threshold of pion
production in $p\gamma$ collisions the neutrino flux has a low-energy
suppression. The low-energy part of neutrino spectrum is produced in
$pp$ collisions of accelerated protons with a gas inside the cavity and
in the SMS interiors. Reflecting from the fireball, the cavity gas
penetrates into the fireball at the Larmor radius distance $d \sim r_L$
for a particle with energy $E' \sim \Gamma m_p$. In this region
accelerated protons interact with an accumulated gas, producing pions
and neutrinos. However, below we shall consider more efficient
mechanisms for production of low-energy neutrinos.

We assume, as it is usually used in the GRB calculations, the
equipartition of total energy distribution in the form of UHE protons
$W_p$, neutrinos $W_{\nu}$, GRB photons $W_{\rm GRB}$ and the energy
transferred to the shock in SMS. Then for neutrino luminosity during
the maximum activity of the considered central source we have $L_{\nu}
\sim (1/4)L_{\rm max}$ with $L_{\rm max}$ given by (\ref{Lmax}). Though
neutrino spectrum is presented by two components, for simplicity of
further estimates we assume the universal spectrum $\propto 1/E^2$.
Normalizing this spectrum to $L_{\nu}$, we obtain for the neutrino
generation rate
\begin{equation} \label{nu-rate}
 Q_{\nu}= \frac{1}{4}\frac{L_{\rm max}}
 {E^2\ln(E_{\nu,\rm max}/E_{\nu,\rm min})}\,.
\end{equation}
Taking into account the neutrino oscillation, one obtains
$\nu_{\mu}+\bar{\nu}_{\mu}$ neutrino flux at a distance $r$ from the
source $F_{\nu_{\mu}+\bar{\nu}_{\mu}}(E)=(1/3)Q_{\nu}(E)/4\pi r^2$.

We shall discuss here only the detection of TeV neutrinos from the
considered hidden source by a future 1~km$^2$ ice or underwater
detector. The fluxes of UHE neutrinos up $10^{19}$~eV are most probably
also detectable, but they need the knowledge of the details of neutrino
detection. This task is beyond the scope of this paper.

Crossing the Earth, neutrinos with energies $E>0.1-1$~TeV produce deep
underground an equilibrium flux of HE muons, which can be calculated
\citep{nu90} as
\begin{equation} \label{Fmu}
  F_{\mu}(>E)=\frac{\sigma_0 N_A}{b_{\mu}}Y_{\mu}(E_{\mu})
  \frac{L_{\rm max}}{12\zeta E_{\mu}} \frac{1}{4\pi r^2}\,,
\end{equation}
where $\sigma_0=1\times10^{-34}$~cm$^2$ is the normalization
cross-section, $N_A=6\times10^{23}$ is the Avogadro number,
$b_{\mu}=4\times10^{-6}$~cm$^2$/g is the rate of muon energy losses,
$\zeta=\ln E_{\rm max}/E_{\rm min}$, and $Y_{\mu}(E)$ is the integral
muon moment of $\nu_{\mu}N$ interaction \citep{book,nu90}. The most
effective energy of muon detection is $E_{\mu}\geq 0.1 - 1$~TeV,
because these muons cross the large part of 1~km$^3$ detector, and
their directions can be reliably measured.  The number of muons
$N(\nu_{\mu})=F_{\mu}S t_{\rm ev}$ with $E_{\mu} \geq 1$~TeV detected
during the burst duration $t_{\rm ev}= 31.3$~days by the underwater
detector with an effective area $S$ at distance $r$ from the source is
given by
\begin{equation} \label{Number}
  N(\nu_{\mu})\simeq 56\;
  \frac{L_{\rm max}}{2.8\times 10^{49}\mbox{ erg s}^{-1}}\;
  \frac{S}{1\mbox{ km}^{2}}
  \left(\!\frac{r}{10^3\mbox{ Mpc}}\right)^{\!-2}\!\!\!.
\end{equation}
The detected muons with $E_{\mu} \geq 1$~TeV show the direction to the
source with an accuracy $2^{\circ} - 3^{\circ}$.

The other acceleration mechanism due to the external ultra-relativistic
shocks, \cite{Vietri95}, provides mostly $pp$ neutrino flux with
energies below 1~TeV. Consider the shock which is produced when a
fireball hits the inner surface of SMS. The shock propagates in
ultra-relativistic regime at distance of order of the Sedov length,
before it overtakes the slow main shock and merges with it. The gas
density in this region rises from that of the cavity $n_c \sim 4\times
10^{11}$~cm$^{-3}$ to maximum density $n_{\rm max} \simeq 54.2 n_{\rm
SMS}$, with $n_{\rm SMS} \simeq 3\times 10^{17}$~cm$^{-3}$ given by
(\ref{rhosms}). The high-energy proton initiates a nuclear cascade due
to pion production. The energy dissipates from the cascade through
cascade-pion decays. The decays of neutral pions initiate the
electro-magnetic cascade and the released energy contributes to the
shock and heating of SMS. The  column density of SMS, $X_{\rm
SMS}\simeq 2.3 \times 10^9M_8^{3/5}$~g~cm$^{-2}$, is large enough to
absorb all produced particles except the secondary neutrinos. Neutrinos
are produced in decays of the charged pions. The critical Lorentz
factor of the decaying charged pion is determined by condition
$\Gamma_{\rm c}\sim(\sigma_{\pi N} \bar{n}\tau_{\pi}c)^{-1} \sim
5\times10^{4}$, where $\sigma_{\pi N} \simeq 25$~mb is $\pi N$
cross-section and $\tau_{\pi}=2.6\times 10^{-8}$~s is the charged pion
lifetime, and for the mean gas density we used $\bar{n} \sim
1\times10^{18}$~cm$^{-3}$, larger than $n_{\rm SMS}$ and smaller than
$n_{\rm max}$.  As a result we obtain for the maximum energy of the
secondary neutrinos $E_{\nu,\rm max} \sim 1$~TeV.

Acceleration by ultra-relativistic shock at the inner boundary of SMS
proceeds in a specific way because of the large density of matter there
and large magnetic field. Consider one of the fireballs hitting the
inner boundary of SMS. Propagating through the dense gas the shock,
running ahead of the fireball, accelerates the protons. A proton from
the SSM interior (upstream region), which makes its first downstream -
upstream - downstream (d-u-d) cycle, i.e. crossing the shock and
scattering upstream back to downstream, increases its energy by a
factor $\Gamma_{\rm sh}^2$. In the consequent d-u-d cycles the energy
gain is only about 2, \cite{GaAch}, (see also \cite{LeRe} for a recent
work). The maximum acceleration energy is determined by the number of
completed d-u-d cycles and is given by $t_a \sim t_{\rm int}$
condition, where $t_a$ and  $t_{\rm int}$ are acceleration and
interaction time, respectively. They are given by $t_a \sim
E/ecB\Gamma_{\rm sh}$, \cite{GaAch}, and $t_{\rm int} \sim
1/c\sigma_{pp}n_{\rm SMS}$. Acceleration time is short because of very
large magnetic field accumulated at the inner boundary of the shock.
The magnetic field in a fireball, $B_{\rm eq}$, given by
Eq.~(\ref{B_eq}) reaches $4\times 10^6$~G, when it hits the SMS. The
each successive fireball adds the magnetic flux, and though not being
summed up coherently, magnetic field increases. For $B \sim 4\times
10^6$~G the maximum energy $E_{\rm max}\sim 1\times10^{17}$~eV.
Fortunately, this energy, being unreliably estimated, is not important
for our calculations of neutrino flux, because all high-energy protons
loose energy by producing nuclear-electromagnetic cascade.

The proton with an energy $E_p$ transfers fraction of energy $f(E_p)$
to neutrinos through decays of cascading charged pions, $\Sigma
E_{\nu}=f(E_p)E_p$. The fraction f=1/2 at $E_p \leq 10^{13}$~eV and
correspondingly $f \sim 0.05$ at $E_p \sim 10^{17}$~eV, because in the
cascading process all neutral pions decay. For the total energies of
protons and neutrinos $W_{\nu}=\bar{f}W_p$, where $\bar{f}$ is the
value averaged over spectrum of accelerated protons. We will assume
$\bar{f} \sim 0.1$. It is easy to see that neutrino signal in the
detector is determined by $W_{\nu}$. It occurs because the
cross-section $\nu N$ at the energies of interest is
$\sigma(E)=\sigma_0 E$, with $\sigma_0=0.8\times 10^{-38}$~cm$^2$/GeV.
The total number of neutrino events in 1~km$^3$ detector is given by
\begin{equation} \label{nu-tot}
N_{\rm tot}=\frac{N_N\sigma_0}{4\pi r^2}\int N_{\nu}(E)E\,dE =
\frac{N_N\sigma_0W_{\nu}}{4\pi r^2}\,,
\end{equation}
where the number of target nucleons is $N_N = 6\times 10^{38}$ and
$W_{\nu} \sim 0.1 W_p$. For $r=1$~Gpc and $W_p=(1/4)W_{\rm tot}=
1.9\times10^{55}$~erg, it results in $N_{\rm tot}=47$ of  low-energy
($E_{\nu} < 1$~TeV) neutrino events in 1~km$^3$ detector. The neutrinos
produced in the SMS by high-energy protons, accelerated in the inner
shocks, double this number of neutrino events.

On may wonder about the influence of adiabatic energy losses of
accelerated protons on neutrino fluxes, estimated above. The role of
adiabatic energy losses has been critically discussed by \cite{Rach}
(see also reply by \cite{wax-repl}. For the calculations above, the
adiabatic energy losses are irrelevant, because in all cases neutrinos
are produced {\em in situ} with the acceleration of particles.

How many these hidden sources are expected? A further (after neutrino
burst) evolution of SMS with a mass $M \sim 10^8 M_{\odot}$ inevitably
results in formation of a massive black hole and the AGN activity.
Considering the described hidden source as a ``pre-AGN'' stage one can
estimate an expected number of hidden sources inside the cosmological
horizon. Since a hidden source and AGN are considered as the different
stages of the same evolving galactic nucleus, the probability to have a
hidden source in the set of AGN is given by the ratio of durations of
the corresponding evolution phases, $t_{\rm ev}/t_{\rm AGN}$, where
$t_{\rm AGN}$ is a duration of AGN phase. Then the number of hidden
sources within the horizon is given by
\begin{equation}
\label{totnum}
 N_{\rm HS} \sim
 \frac{4}{3}\pi (3ct_0)^3 n_{\rm AGN}~t_{\rm s}/t_{\rm AGN}\,,
\end{equation}
where $(4\pi/3)(3ct_0)^3$ is the cosmological volume inside the horizon
$ct_0$ and $n_{\rm AGN}$ is the number density of AGN. The estimates
for $n_{\rm AGN}$ and $t_{\rm AGN}$ taken for different populations of
AGN results in $N_{\rm HS} \sim 1 - 10$.

Can the hidden sources be observed in electromagnetic radiations? X-ray
and gamma-ray radiations are fully absorbed due to tremendous thickness
of SMS, $X_{\rm SMS} \sim 2 \times10^{9}$~gcm$^{-2}$. However the
thermaized radiation can propagate diffusively and reach the the outer
surface of SMS. This process has been considered in \cite{ber01} for
the case of a massive shell. In case of SMS, the diffusion is much
slower and the thermalized radiation reach the outer surface much
later, and thus will be emitted with a lower luminosity. The diffusion
coefficient $D \sim c l_{\rm d}$, where a diffusion length is $l_{\rm
d} = 1/(\sigma_{\rm T}n_{\rm SMS})$, and $\sigma_{\rm T}$ is the
Thompson cross-section. A mean time of the radiation diffusion through
SMS is very long, $t_d  \sim R_{\rm SMS}^2/D \sim
4\times10^6M_8^{4/5}$~yrs, in comparison with a lifetime of the NS
cluster inside the SMS. Since the dispersion of the diffusion time
distribution $\sigma \sim t_d$, the surface luminosity of the central
source is $L_{\rm bb} \sim E_{\rm tot}/\sigma \sim L_{\rm max}t_{\rm
ev}/t_d\sim5\times 10^{41}$~erg/s and hence is negligible in comparison
with the Eddington luminosity of the host SMS.

\section{Conclusions}

In this paper we studied the natural scenario for massive black hole
formation in the process of dynamical evolution of a compact stellar
cluster. Our contribution to this scenario is based on the observation
that an intermediate SMS can be formed with a subsystem of neutron
stars submerged in it. The evolution of neutron star cluster results in
its contraction and merging of neutron stars at their collisions. Both
these processes have been studied before and rather well known. The
merging of neutron stars can produce quasi-spherical fireballs, where
protons are accelerated to high energies, and neutrinos are produced.

Described in a more detailed way our scenario looks as follows. The
dynamical evolution of a central stellar cluster in the galactic
nucleus results in the collision destruction of normal stars and
formation of massive gas envelope \citep{ber01} or SMS (this paper).
The neutron stars and stellar mass black holes survive through this
stage and form self-gravitating subsystem deep inside the SMS. The
lifetime of this compact neutron star cluster is very short. The
multiple fireballs after neutron star collisions in this cluster dig
out a rarefied cavity inside the SMS. The protons are accelerated in
the fireballs and in the cavity and produce neutrinos in collisions
with an ambient photons and gas. The active phase of high luminosity of
the central source has a duration $t \sim 0.1 - 1$~yr. During the
neutrino burst the energy of the same order as a total neutrino energy
${\cal E}_{\nu}$ is released also in the form of high-energy electrons
and photons. All high-energy particles, except neutrinos, are
completely absorbed in the SMS, which column density can reach $X \sim
10^9$~g/cm$^2$. Thus the considered object is a very powerful hidden
neutrino source. It is demonstrated that that shocks after numerous
fireballs are absorbed inside the SMS and cannot destroy it. The system
remains gravitationally bound and gravitational collapse of the compact
neutron star cluster produces a seed massive black hole inside a host
SMS. Long after the neutrino burst, the described hidden neutrino
source is to be seen as a bright AGN. As precursors of most powerful
AGN, these hidden sources are expected to be at the same redshifts as
AGNs. The number of the hidden neutrino sources can be estimated as
$\sim1 -10$ in the visible universe.

The hidden neutrino source is detectable by future 1~km$^3$ detector
with a number of TeV muons up to $50 - 100$ per source. The predicted
number of low-energy ($E_{\nu} < 1$~TeV) neutrino interactions is $\sim
100$. The highest energy neutrinos with $E > 10^{15}$~eV are probably
also detectable.

\begin{acknowledgements}
We acknowledge the useful discussions with Bohdan Hnatyk and Yury
Eroshenko.  We are grateful to anonymous referee for valuable comments.
This work was supported in part by the Russian Foundation for Basic
Research grants 04-02-16757 and 06-02-16342, and the Russian Ministry
of Science grant 1782.2003.2.
\end{acknowledgements}


\begin{thebibliography}{99}

\bibitem[Benz and Hills(1992)]{ben92} Benz, W., and Hills, J. G. 1992,
    \apj, 389, 546

\bibitem[Blandford and McKee(1976)]{bla76} Blandford, R. D., and
    McKee, C. F. 1976, Phys. Fluids, 19, 1130

\bibitem[Berezinsky and Smirnov(1975)]{BeSm} Berezinsky, V. S., and
Smirnov A. Yu. 1975, Ap. Sp. Sci., 32, 461

\bibitem[Berezinsky(1990)]{nu90} Berezinsky, V. 1990, Nucl. Phys. B
    (Proc. Suppl.), 19, 375

\bibitem[Berezinsky et al.(1990)]{book} Berezinsky, V. S.,
    Bulanov, S. V., Dogiel, V. A., Ginzburg,  V. L. and
    Ptu\-skin, V. S. 1990,  Astrophysics of Cosmic Rays,
    North-Holland, Amsterdam

\bibitem[Berezinsky and Dokuchaev(2001)]{ber01} Berezinsky, V. S., and
    Dokuchaev, V. I. 2001, Astropart. Phys. 15, 87

\bibitem[Berezinsky and Ginzburg(1981)]{ber81} Berezinsky, V. S., and
    Ginzburg, V.L. 1981,  194, 3

\bibitem[Berezinsky and Prilutsky(1987)]{BePr} Berezinsky, V. S., and
    Prilutsky, O. F.  \aap, 66, 325

\bibitem[Chandrasekhar(1943)]{cha43} Chandrasekhar, S. 1943, \apj,
    97, 255

\bibitem[Colgate(1967)]{col67} Colgate, S. A. 1967, \apj, 150, 163

\bibitem[Cohn(1980)]{coh80} Cohn, H. 1980, \apj, 242, 765

\bibitem[Dokuchaev(1991)]{dok91} Dokuchaev, V. I. 1991, \mnras, 251, 564

\bibitem[Dokuchaev et al.(1998)]{dok98} Dokuchaev, V. I.,
    Eroshenko, Yu.N., and Ozernoy, L. M. 1998, \apj, 502, 192

\bibitem[Dokuchaev(2002)]{dok02} Dokuchaev, V. I. 2002, \aap, 395, 1023

\bibitem[Gallant and Achterberg(1999)]{GaAch} Gallant Y. A., and
    Achterberg A. 1999, \mnras, 305, L6

\bibitem[Lai et al.(1993)]{lai93} Lai, D., Rasio, F. A., and
    Shapiro, S. L. 1993, \apj, 412, 593

\bibitem[Landau and Lifshitz(1959)]{ll59} Landau, L. D. and
    Lifshitz, E. M. 1959, Fluid Mechanics, Addison-Wesley Reading, Mass.,
    Chapters X, \S~106 and XV

\bibitem[Lemoine and Revenu (2005)]{LeRe} Lemoine~M. and Revenue~B. 2005,
    {\tt arXiv: astro-ph/0510522}

\bibitem[New and Shapiro(2001)]{new01} New, K. C. B., and Shapiro, S. L.
    2001, \apj, 548, 439

\bibitem[Ostriker(1999)]{ost99} Ostriker,  E. C. 1999, \apj, 513, 252

\bibitem[Ostriker \& McKee(1995)]{ost88} Ostriker J. P., and
    McKee C. F. 1988, Rev. Mod. Phys., 61, 1

\bibitem[Piran(2000)]{pir00} Piran, T. 2000, \physrep, 333, 529

\bibitem[Quinlan and Shapiro(1987)]{qui87} Quinlan, C. D., and
    Shapiro, S. L. 1987, \apj, 321, 199

\bibitem[Quinlan and Shapiro(1990)]{qui90} Quinlan, C. D., and
    Shapiro, S. L. 1990, \apj, 356, 483

\bibitem[Rachen and Meszaros(1998)]{Rach} Rachen J. P. and
    Meszaros P., 1998, \prd,  58, 123005, and
    {\tt arXiv:astro-ph/9811266}

\bibitem[Rees(1984)]{ree84} Rees, M. J. 1984, \araa, 22, 471

\bibitem[Sanders(1970)]{san70} Sanders, R. H. 1970, \apj, 162, 791

\bibitem[Saslaw(1987)]{Saslaw} Saslaw, W. C. 1987, Gravitational
    physics of stellar and galactic systems. Cambridge Univ.
    Press: Cambridge

\bibitem[Shapiro and Teukolsky(1983)]{sha83} Shapiro, S.  L.,
and Teukolsky, S.  A.  1983, Black Holes, White Dwarfs and Neutron
Stars (New-York:  Willey)

\bibitem[Spitzer(1987)]{Spitzer} Spitzer, L. 1987, Dynamical Evolution
    of Globular Clus\-ters. Princeton Univ. Press: Princeton

\bibitem[Spitzer(1971)]{spi71} Spitzer, L. 1971, Galactic Nuclei,
    D.~Q'Connel, North Holland, Amsterdam, 443

\bibitem[Spitzer and Saslaw(1966)]{spi66} Spitzer, L., and Saslaw, W.C.
    1966, \apj, 143, 400

\bibitem[Stecker et al.(1991)]{Ste91} Stecker, F. W., Done, C.,
    Salamon, M. H., and Sommers, P. 1991, \prl, 66,  2697

\bibitem[Vietri(1995)]{Vietri95} Vietri, M. \apj, 453, 883

\bibitem[Waxman(1995)]{wax95} Waxman, E., 1995, \prl, 75, 386

\bibitem[Waxman(2001)]{wax01} Waxman, E., 2001,  Lect. Notes Phys.,
     576, 122 {\tt arXiv: astro-ph/0103186}

\bibitem[Waxman(2004)]{wax-repl} Waxman E., 2004. \apj 606, 988

\bibitem[Waxman and Bahcall(1997)]{WaBa97} Waxman, E. and Bahcall, J.
\prl, 78, 2292

\bibitem[Thorne and Zytkow(1977)]{ThZy} Thorne, K. S. and Zytkow, A.N.
    \apj, 212, 832

\bibitem[Zel'dovich and Novikov(1971)]{zel71} Zel'dovich, Ya.  B.  and
Novikov I.  D.  1971, Relativistic Astrophysics, Vol.~1 (Chicago: Univ.
Chicago Press)

\bibitem[Zel'dovich and Poduretz(1965)]{zp65} Zel'dovich, Ya. B. and
Poduretz, M. A. 1965, Soviet Astr., 9, 742

\end{thebibliography}
\end{document}